\documentclass[conference, 11pt]{IEEEtran}
\usepackage{iabproject}
\usepackage{times}
\usepackage{graphicx}
\usepackage{amsmath, amssymb,latexsym}
\usepackage{url}

\wintitle[Spring 2018]{Edge Cloud System Evaluation 
}

\author{
\IEEEauthorblockN{Sumit Maheshwari and Dipankar Raychaudhuri}
\IEEEauthorblockA{WINLAB, Rutgers University, North Brunswick, NJ, USA}
\{sumitm, ray\}@winlab.rutgers.edu
}

\begin{document}
\maketitle
\begin{abstract}
Real-time applications in the next generation networks often rely upon offloading the computational task to a \textit{nearby} server to achieve ultra-low latency. Augmented reality applications for instance have strict latency requirements which can be fulfilled by an interplay between cloud and edge servers. In this work, we study the impact of load on a hybrid edge cloud system. The resource distribution between central cloud and edge affects the capacity of the network. Optimizing delay and capacity constraints of this hybrid network is similar to maximum cardinal bin packing problem which is NP-hard. We design a simulation framework using a city-scale access point dataset to propose an enhanced capacity edge cloud network while answering following questions: (a) how much load an edge cloud network can support without affecting the performance of an application, (b) how is application delay-constraint limit affects the capacity of the network, (c) what is the impact of load and resource distribution on goodput, (d) under what circumstances, cloud can perform better than edge network and (e) what is the impact of inter-edge networking bandwidth on the system capacity. An evaluation system and model is developed to analyze the tradeoffs of different edge cloud deployments and results are shown to support the claims. 
\end{abstract}
\section{Introduction}
Emerging low-latency class of applications has pushed network components close to user as cloud or edge servers. On one hand, there is an exponential growth in the number of devices being connected to the Internet while on the other hand, sophisticated software such as augmented and virtual reality (AR/VR), running on those devices are able to consume data in the order of milliseconds. In this work, we attempt to find if these low-latency applications can be supported by edge clouds in cooperation with central cloud when resources such as bandwidth and computing power are limited and if so, what must be the right resource distribution to support an application with service quality guarantee. At the application front, surge in the demand of ultra-low latency applications for instance annotation based assistance using Augmented Reality (AR) pose strict bounds on compute as well as network latency. User Equipment (UE) running AR applications, in general, sends a continuous video or audio stream (along with metadata such as user velocity, location) to the server or cloud for processing and a calculated output is received by the UE based upon the application type which overturns typical HTTP traffic which is heavy-tailed, read-mostly model wherein users are the data consumers and server is the generator. The paradigm shift of higher upload traffic for certain applications than the download traffic is visible in AR/VR services where the UE receives only contextual outputs from server (cloud or edge) while server continuously processes each and every frame of the video stream thus requiring higher computation ability close to user in order to provide output in \textit{true} real-time. 
\vspace{-6mm}
\section{Approach}
\label{sec:app}
In order to understand real-time applications’ performance over edge cloud network, we developed a tool for city scale network simulations. We use Chicago city which is the third most populous city in US as a test-case considering locations of 11,00 Wi-Fi APs as shown in Fig. 1. Each AP is equipped with an edge server with a configurable compute resource capacity. A rack has limited capacity to support users for their computational requirements. For instance, an AR application requires computation to process video/image stream and receive the response back from the server. Therefore, upgrading an edge means deploying more machines to enhance capacity or simply updating the software. The central cloud server is placed at Oregon which again has a configurable capacity. For a system without edge clouds, the central cloud should be able to support all the users. In our simulation, we increase the resource density of already deployed edge clouds by removing compute resources from the central cloud. This method keeps the compute cost for the whole system unchanged. Therefore, the total system capacity including the central cloud and edge is constant in terms of compute.

\begin{figure}[t!]
\begin{center}
\vspace{-1mm}
\includegraphics[width=0.5\linewidth]{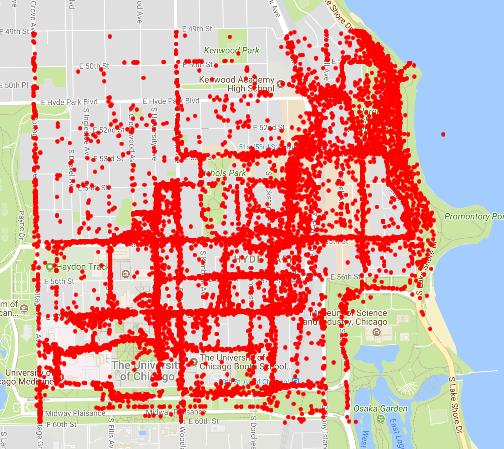}
\vspace{-2mm}
\caption{Access Point Distribution at the Chicago City}
\vspace{-10mm}
\label{fig:ap}
\end{center}
\end{figure}

Application is modeled using a four tuple $<V,G,S,L>$. V: compute task, G: geographical block, S: availability of edge in a block and L: tolerable latency for an application. Compute at edge server and at the cloud is modeled using M/M/C queueing. Latency of cloud is given as ${{L}_{cloud}}=(\alpha +\delta )*{{D}_{\min (UE, APs)}}+(\beta +\gamma)*{{D}_{AP-cloud}}+{{d}_{node}}$, where distances from UE to AP and AP to cloud are considered with their respective factors based upon ping statistics and available bandwidth. Latency at edge is given by ${{L}_{edge}}=(\alpha +\delta )*{{D}_{\min (UE, APs)}}+{{d}_{node}}+{d}_{s}$, where $d_{node}$ is the sum of processing and queueing latency, and $d_{s}$ is the switching latency from one edge to another available edge. In the baseline approach, an edge selects an available neighbor while in the proposed approach, ECON, an edge with more available resources is chosen irrespective of geographical closeness with the UE. An edge or cloud is “usable” for a request $i$ if the latency for the user running an application a is below the latency threshold for given application. “Delay-constraint (\%)” of an edge-cloud system is defined as the number of requests out of hundred served below the application response time threshold. For a given threshold, Delay-constraint can also be interpreted as system capacity. For instance, a Delay-constraint of 10\% for a 15 ms threshold implies that system can accommodate only 10\% of the total requests and 90\% requests will only consume resources to lower the goodput. This means for 90\% of the requests, the assigned edge resources are “not usable”. The system is analyzed for different resource distribution at edge and at the cloud, with different AP to cloud and inter-edge bandwidth, and various system loads. \\
Figure 2 compares the goodput of ECON vs. Baseline for different load and resource distribution. For a cloud-favored resource case, ECON optimizes the capacity of edge-cloud network by choosing any available edge as comapred to the Baseline. For Edge-favored resource case, ECON and Baseline have similar performance as finding a neighbor is equally good as finding the available edge \cite{sumit}.

\begin{figure}[t!]
\begin{center}
\vspace{-4mm}
\includegraphics[width=1\linewidth]{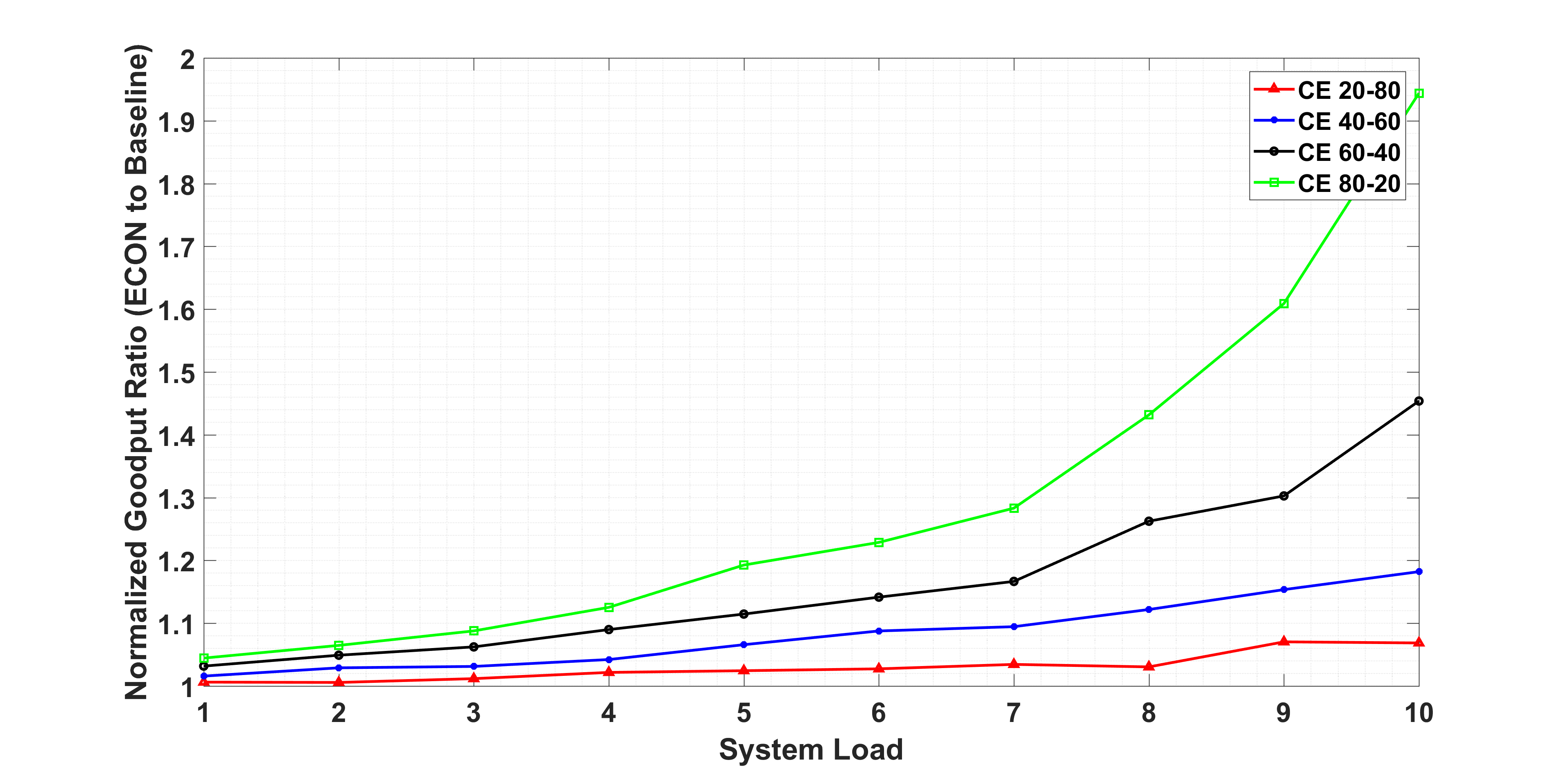}
\vspace{-7mm}
\caption{Impact of Load on Goodput Ratio of ECON and Baseline in an Edge Cloud System for Real-time Applications for Different Resource Distribution}
\vspace{-5mm}
\label{fig:ap}
\end{center}
\end{figure}
\vspace{-6mm}
\section{Conclusion}
\label{sec:conclusion}
In this work, a distributed edge selection (baseline) scheme is compared with a generalized decision (ECON) approach for various system load, edge-cloud resource distribution, inter-edge bandwidth and AP-cloud bandwidth parameters. A mechanism to choose operating point on a network with respect to the load is provided for instance when Load=7, a cloud-only system outperforms an edge-only 1 Gbps inter-edge bandwidth system. Delay-constraint enables an application service provider to choose the combination of edge and cloud resources with available bandwidth to provide best possible quality. The goodput analysis highlights the optimization avenues for a hybrid edge cloud system as for Load=8, Cloud-Edge 60\%-40\% resource distributed system, we can achieve at least 20\% more goodput than the baseline. We show that for higher inter-edge bandwidth and edge-favored resources case, a distributed edge selection is equally good as finding a central optimal whereas for cloud-favored resources and no bandwidth constraints, ECON provides lower average response time. Our study depicts that adding capacity to an existing edge resource without increasing internetwork bandwidth may actually increase network-wide congestion and can reduce capacity of the network. Future work includes analyzing the impact of mobility on the system capacity and edge placement. 

\section*{Acknowledgement}
Research supported under NSF Future Internet Architecture - Next Phase (FIA-NP) Award CNS-134529

\bibliography{projref}

\end{document}